\documentclass{appolb}
\usepackage{epsfig}
\usepackage{amsmath,amssymb}

\newcommand{\beqn}{\begin{equation}}
\newcommand{\eeqn}{\end{equation}}
\newcommand{\req}[1]{Eq.\,(\ref{#1})}

\newcommand{\Tmm}{{\cal T}}
\newcommand{\Fumn}{F^{\mu\nu}}
\newcommand{\Humn}{H^{\mu\nu}}
\newcommand{\Fmn}{F_{\mu\nu}}
\newcommand{\dFumn}{\widetilde{F}^{\mu\nu}}
\newcommand{\Tumn}{T^{\mu\nu}}
\newcommand{\gumn}{g^{\mu\nu}}
\newcommand{\Veff}{V_{\rm eff}}

\begin{document}
\title{Nonlinear Electromagnetic Forces in Astrophysics%
\thanks{Presented by LL at 52 Krak\'ow School of Theoretical Physics: Astroparticle Physics in the LHC Era, Zakopane, May 19-27, 2012 }
}
\author{Lance Labun$^{1,2}$ and Jan Rafelski$^1$
\address{$^1$~Department of Physics, The University of Arizona, Tucson, 85721 USA\\
$^2$~Leung Center for Cosmology and Particle Astrophysics,\\National Taiwan University, Taipei 10617, Taiwan}
}
\maketitle
\begin{abstract}
Electromagnetism becomes a nonlinear theory having (effective) photon-photon interactions due at least to electron-positron fluctuations in the vacuum.  We discuss the consequences of the nonlinearity for the force felt by a charge probe particle, and compare the impact of Euler-Kockel QED effective nonlinearity to the possibility of  Born-Infeld-type nonlinearity.  
\end{abstract}
\PACS{03.50.De,12.20.-m,95.30.-k}   

\section{Lorentz force and Quantum Vacuum as polarizable medium}
Given the weakness of gravity, a natural question is, if nonlinear electromagnetic interactions can on stellar scales lead to transport of   matter in ways differing from the Lorentz force expectation.  This question is especially pertinent for stellar objects having extreme magnetic fields (magnetars), and we investigate here in a covariant formulation additional forces induced in extreme magnetic fields. 

The Lorentz force predicts that a charged particle moving in a magnetic field of arbitrary strength will experience a force normal to both the field direction and the direction of motion. This pivotal property decides the fate of charged particles in the magnetic field-filled Universe. Even a very small field suffices to determine the particle dynamics when acting over a distance large compared to the natural microscopic particle scale, and does so as a function of their velocity. Electric fields play a subdominant role on macroscopic scale since their presence requires a separation or imbalance of charge which normally cannot be maintained for a long time or/and in a large volume. 

The  Lorentz force has not been well tested on a macroscopic scale. It could be modified even on planetary scale, considering the material constants of a planetary space plasma, the dielectric polarizibility $\epsilon$ and magnetic permeability $\mu$. The question we wish to address here is if, and  under what conditions analogue physical effects originate in the empty space vacuum polarizibility due to vacuum fluctuations and/or a natural nonlinearity of electromagnetism. These effects are normally studied in atomic environments, and in comparison to the dominant Coulomb force are exceedingly small. Precision atomic experiments in part compensates this smallness. However as we will argue here the astrophysical environment may offer an alternate approach to this physics domain of  nonlinear electromagnetism. 

\begin{figure}
\centerline{\includegraphics[width=0.3\textwidth]{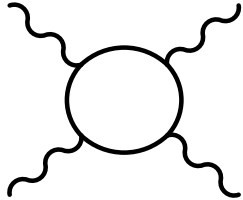}}
\caption{Feynman diagram expressing effective field-field interaction going through electron-positron fluctuation.\label{FFdiag} }
\end{figure}

The most famous example of modification of the classical electromagnetic theory is the anomalous (since absent unless quantum fluctuations are considered) light-light scattering phenomenon.  Early in the development of quantum electrodynamics, Euler and Kockel~\cite{Euler:1935zz} (and Heisenberg and Euler~\cite{Heisenberg:1935qt}, see below) recognized that electron-positron fluctuations in the vacuum would generate an effective self-interaction for the electromagnetic field.  This effect is visualized in the language of Feynman diagrams as the four electromagnetic field legs attached to an electron loop, see Figure~\ref{FFdiag}.

One may also view this phenomenon as the presence of the external field $F^{\mu\nu}$ polarizing the electron fluctuations in the vacuum.  This polarization of vacuum fluctuations is analogous to the modification  of electromagnetic properties in a material body, e.g. plasma, but with several important differences of fundamental origin: 
\begin{enumerate}
\item 
The QED vacuum is Lorentz invariant, whereas a material body has a preferred reference frame in which the medium is at rest.  This means that, among other things, the framework discussed here and the physics it represents provide a consistent Lorentz invariant theory of nonlinear electromagnetism with sources.
\item
In QED, the linear polarization 
$$\Fumn\Fmn\to (1+(\alpha/3\pi)\ln(-q^2/m_e^2))\Fumn\Fmn$$ 
is divergent and is absorbed in the renormalization of charge.  In a material body, the linear polarization $\vec E\to \vec D=(1+\overset{\leftrightarrow}{\varepsilon}_0)\vec E$ is typically finite.
\item  
The (consequences of QED) polarization we discuss arise from the nonlinear response, such as visualized in Fig.\ref{FFdiag}, and can be presented in terms of a dielectric tensor $\overset{\leftrightarrow}{\varepsilon}$ that is necessarily a covariant function of the applied quasi-constant field. 
\end{enumerate}

Heisenberg and Euler showed how to calculate the polarization effect to all orders in the applied (external) field~\cite{Heisenberg:1935qt}.  In so doing, they obtained the first non-perturbative result in quantum field theory, which has been studied and generalized extensively in the intervening years; for a survey of this development we refer the reader to the recent review~\cite{Dunne:2004nc}.  The outcome  is an effective potential $\Veff$ for the electromagnetic field, providing the paradigmatic example of effective field theory, in this case arising by integrating out a ``heavy'' electron in the low energy ($\omega\ll m_e$) limit of large scale (in space and time) quasi constant fields.  Their result contains the Euler and Kockel result seen in Fig.\ref{FFdiag} at the lowest finite non-vanishing order (after renormalization) expansion in the fine structure constant $\alpha$.

However, the full Heisenberg-Euler nonlinearity of electromagnetism is important, considering certain compact stellar objects, `magnetars', which are now widely expected to be harboring magnetic fields of equal or greater magnitude to the so-called ``critical'' strength of QED~\cite{TML:1993,Mereghetti} 
\beqn\label{Bc}
|\vec B_c|=\frac{m^2c^2}{e\hbar}=4.41\:10^{13}~{\rm G}.
\eeqn
Near this field strength,  all  higher order terms in the effective potential become of the same order as is the lowest Euler-Kockel term, see figure~\ref{FFdiag}. 

We will also explore in this report  the effective force in a natural nonlinear extension of Maxwell linear electromagnetism, that is the limiting field Born-Infeld electromagnetism~\cite{BornInfeld}.  Considering microscopic physics, we show that  new constraints on the Born-Infeld theory require first understanding the dominant QED (Heisenberg-Euler) effects at the percent level.

In the following, we will recall briefly the framework of classical nonlinear electrodynamics and how to derive the dynamics of charged particles and fields.  We will discuss why it is necessary to consider energy-momentum conservation explicitly in deriving the effective electromagnetic force and obtain a general formula for the leading terms in the force valid in the leading order in the expansion of the nonlinear effective potential. This corrects our earlier presentation in theoretical detail~\cite{Labun:2008gm}, though the magnitude of the effects remains.

\section{Classical Dynamics}
The Lorentz equation of motion of  particles, and Maxwell field equations, are derived by minimizing the total action. 
\beqn\label{EMaction}
I_{\rm tot}=\int d^4x{\cal L}_{\rm tot},\quad  {\cal L}_{\rm tot}=-{\cal S}+\Veff+j^{\mu}A_{\mu}+m\frac{1}{2}(u^{\mu}u_{\mu}-1) .
\eeqn 
Here the Maxwell electromagnetic action $-{\cal S}=(\vec E^2-\vec B^2)/2$ is supplemented by an effective potential, such as the Heisenberg-Euler $\Veff$.  The Maxwell electromagnetic field equations follow from the Euler-Lagrange equations, varying \req{EMaction} with respect to $A_{\mu}$.

Note that the Lorentz force emerges as consequence of variation of the {\em path} a charged particle takes in the field-filled space. However, a more unified view of  paths with  fields arises when we consider the  Lorentz force as emerging from conservation of energy-momentum tensor. The Feynman path integral formulation of QED offers yet greater unification of both dynamical elements (fields and paths) but this discussion goes well beyond the scope of this report.

At first, we will not specify what $\Veff$ is, in order to provide a general derivation and discussion of the consequences of the presence of any $\Veff$.  The third term in \req{EMaction} is the gauge invariant minimal coupling of the electromagnetic potential to the current, and the last term encodes the charged particles' (inertial) mass $m$ and four-velocity $u^{\mu}=\gamma(1,\vec v)$. Note that at this point we do not differentiate between electromagnetic and mechanical mass of a particle: there is inertia in the electromagnetic field that accompanies a charged particle and a redefinition of mass is required. This will be achieved below in \req{TmnintDefn} where we consider the force related to the nonlinearity of the field.

The effective electromagnetic Lagrangian 
\beqn\label{LtotEM}
{\cal L} =-{\cal S}+\Veff
\eeqn
can depend only on the scalar and pseudoscalar Lorentz invariants
\beqn
{\cal S}=\frac{1}{4}\Fumn\Fmn, \quad {\cal P}=\frac{1}{4}\dFumn\Fmn, \quad \dFumn \equiv \frac{1}{2}\varepsilon^{\mu\nu\kappa\lambda}F_{\kappa\lambda}
\eeqn
Varying \req{EMaction} with respect to $A_{\mu}$ then yields
\beqn\label{EMsource}
\partial_{\mu}\Humn = j^{\nu}
\eeqn
where $\Humn$ defined as the displacement field tensor (see for example \S 6 of~\cite{BialynickiQED})
\beqn\label{displacementdefn}
H^{\mu\nu}\equiv -\frac{\partial{\cal L}_{\rm tot}}{\partial\Fmn} 
=\Fumn -\frac{\partial\Veff}{\partial\mathcal{S}}\Fumn		-\frac{\partial\Veff}{\partial\mathcal{P}}\dFumn
\eeqn
When $\Veff\to 0$, note that $\Humn\to\Fumn$, and \req{EMsource} becomes the standard Maxwell equation with source, $\partial_{\mu}\Fumn = j^{\nu}$.

Current conservation means the field tensor is the 4-dimensional curl of a vector potential $A^{\mu}$ and implies the homogeneous Maxwell equation
\beqn\label{homoMaxeqn}
\partial_{\mu}\dFumn=0.
\eeqn

\section{Charged Particles in External Fields}

The dynamics of the charged particle plus field system can be derived from the statement of joint electromagnetic and matter energy-momentum conservation, 
\beqn\label{conservTmn}
\partial_{\mu}(\Tumn_{\rm e.m.}+\Tumn_{\rm matter}) = 0
\eeqn
The matter (last) part of the action \req{EMaction} defines $\Tumn_{\rm matter}$ with the property 
\beqn
\partial_{\mu}\Tumn_{\rm matter}=u^{\mu}\partial_{\mu}(mu^{\nu})=dp^{\mu}/d\tau.
\eeqn

At this point, one might be tempted to identify $-\partial_{\mu}\Tumn_{\rm e.m.}$ as the force applied to the charged probe particle.  However, this calculation yields the standard Lorentz force $j_{\mu}\Fumn$, even in the presence of nonlinear electromagnetism (see Appendix~\ref{app:LF}).  The correct approach to recognize the force is inherent in the work of Born-Infeld~\cite{BornInfeld}, which considers the field energy-momentum of individual particles to be  identified with  their electromagnetic inertia, and hence included in the definition of inertial mass $m$ of the particle sourcing the field.  We extend this also to the case of vacuum fluctuation nonlinearity.

By subtracting the energy-momentum of the separate probe particle and external  field components as they appear in isolation, only the nonlinear interaction part of the electromagnetic field tensor is  retained as sourcing the particle-particle force, see section 4 in Ref.~\cite{RafelskiBI3},
\beqn\label{TmnintDefn} 
\Tumn_{\rm int}=\Tumn_{\rm e.m.}-\Tumn_{\rm p}-\Tumn_{\rm e}.
\eeqn  
The electromagnetic force is exhibited as the divergence of interaction energy-momentum $\Tumn_{\rm int}$. This procedure will show the usual Lorentz force, as well as further contributions due to the nonlinear interaction of the charged particle's electromagnetic field with the external field, see Appendix~\ref{app:deltafderiv}

For a general nonlinear electromagnetic theory the energy-momentum tensor can be written~\cite{Schwinger:1951nm} (see also Eq. (5) of~\cite{Labun:2008qq})
\beqn\label{Tmn-em}
\Tumn_{\rm e.m.} = \varepsilon\Tumn_{\rm Max}+\gumn \Tmm/4,\quad 
\Tumn_{\rm Max}=\gumn{\cal S}-F^{\mu\kappa}F^{\nu}_{\phantom{\nu}\kappa} 
\eeqn
The deviation from the Maxwell energy-momentum tensor $\Tumn_{\rm Max}$
is described by the two Lorentz-scalar functions 
\begin{align}\label{epsdefn}
&\varepsilon=-\frac{\partial(-{\cal S}+\Veff)}{\partial{\cal S}}=1-\frac{\partial\Veff}{\partial{\cal S}},\\
&\Tmm\equiv T^{\mu}_{\mu}=-4\left(\Veff-{\cal S}\frac{\partial\Veff}{\partial{\cal S}}-{\cal P}\frac{\partial\Veff}{\partial{\cal P}}\right) \label{Tmmdefn} 
\end{align}
The energy-momentum trace $\Tmm$ must arise from $\Veff$, because the energy-momentum tensor of the Maxwell theory $\Tumn_{\rm Max}$ is traceless, and it must start at order $e^4$, being related to the presence of (effective) field-field interactions~\cite{Adler:1976zt}, such as displayed in Fig.~\ref{FFdiag}.

\begin{figure}
\centerline{\includegraphics[width=0.3\textwidth]{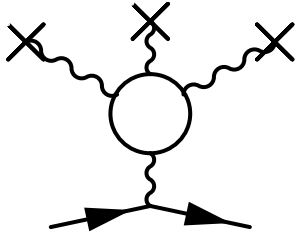}}
\caption{Diagram expressing contribution to the effective force arising from light-by-light scattering in QED.  In the Born-Infeld theory, the loop is to be seen as collapsed to a point.  \label{FFforce} }
\end{figure}

For the case of an external magnetic field, the interaction energy-mo\-mentum $\Tumn_{\rm int}$ is calculated to leading order in Appendix~\ref{app:deltafderiv}.  This depends on three powers of the external field and is order $e^4$, which one can see  in Fig.~\ref{FFdiag} by counting the vertices, each of which comes with a power of $e$.\footnote{We count powers with the QED Heisenberg-Euler $\Veff$ in mind, and we will see that the count is simply decreased by 2 for Born-Infeld theory~\cite{BornInfeld} due to the absence of $\alpha/\pi$ arising from the fermion loop.}  In the case of QED nonlinearity we are thus finding the effect represented schematically in Fig.~\ref{FFforce}.

Taking the divergence of \req{Tmnint},
\beqn\label{cov-f-defn}
\frac{dp^{\mu}}{d\tau}=-\partial_{\nu}\Tumn_{\rm int}
    =-j_{\nu}F_{\rm e}^{\nu\mu}+\delta\! f^{\mu}
\eeqn
For the second equality, the Lorentz force (density) is obtained from the divergence of the Maxwell-Lorentz interaction $\Tumn_{\rm ep}$ \req{Tmnep} and separated.  

The remainder is the sought-for modification
\begin{align}\label{delta-4force}
\delta\! f^{\mu} \simeq~ &
\frac{\partial\varepsilon}{\partial{\cal S}}F^{\rm e}_{\kappa\lambda}H_{\rm p}^{\kappa\lambda}\partial_{\nu}\Tumn_{\rm Max,e}
+ \Tumn_{\rm Max,e}\partial_{\nu}\left(\frac{\partial\varepsilon}{\partial{\cal S}}F^{\rm e}_{\kappa\lambda}H_{\rm p}^{\kappa\lambda}\right)  \\
&-\gumn\partial_{\nu}\left(\frac{1}{4}\frac{\partial\Tmm}{\partial{\cal S}}F^{\rm e}_{\kappa\lambda}H_{\rm p}^{\kappa\lambda}\right) \notag
\end{align}
where the Maxwell energy-momentum tensor $\Tumn_{\rm Max,e}$ and partial derivatives $\partial\varepsilon/\partial{\cal S},\partial\Tmm/\partial{\cal S}$ are evaluated at the external field.

\section{Born-Infeld -- Euler-Kockel Comparison}
The action of Born-Infeld electrodynamics is
\begin{align}\label{LBI}
{\cal L}_{\rm BI}&=({\cal E}_c)^2-({\cal E}_c)^2\sqrt{1+2{\cal S}/{\cal E}_c^2 -{\cal P}^2/{\cal E}_c^4}\\
&=-{\cal S}+\frac{1}{2{\cal E}_c}({\cal S}^2+{\cal P}^2)-\frac{\cal S}{2{\cal E}_c^2}({\cal S}^2+{\cal P}^2)+...
 \notag\end{align}
Born  selected originally the limiting electric field strength, ${\cal E}_c$, such that the inertia of the electromagnetic field of an electron was equal to the electron mass. Appearance of other fundamental charged particles, the quantum-field theory path to mass renormalization, and further mechanisms (such as Higgs) to generate mechanical particle mass reveal that ${\cal E}_c\equiv M^2/e$  is a free parameter, constrained by experiment~\cite{RafelskiBI1}. Limitations of the possible range of values of the associated mass scale  $M$  will be discussed below in comparison with QED.

The expression for the 4-force \req{delta-4force} is given in terms of $\varepsilon$ and $\Tmm$, which we can obtain analytically from \req{LBI}.  The coefficient functions appearing in \req{deltaf0} are 
\begin{align}
\label{BIeps}
\frac{\partial\varepsilon}{\partial{\cal S}}&=-(1+2{\cal S}/{\cal E}_c^2-{\cal P}^2/{\cal E}_c^4)^{-3/2}{\cal E}_c^{-2}=-\varepsilon^3{\cal E}_c^{-2}\\
&\simeq -\frac{e^2}{M^4}+\frac{3e^4{\cal S}}{M^8}-\frac{e^6}{M^{12}}\frac{3}{2}(5{\cal S}^2+{\cal P}^2)+... \notag\\
\label{BITmn}
\frac{1}{4}\frac{\partial\Tmm}{\partial{\cal S}}&=\varepsilon^3({\cal S}/{\cal E}_c^2-{\cal P}^2/{\cal E}_c^4)\\
&\simeq \frac{\cal S}{{\cal E}_c^2}-3\frac{{\cal S}^2}{{\cal E}_c^4}-\frac{{\cal P}^2}{{\cal E}_c^4}... \notag
\end{align}
given also in weak-field expansion in the second line.

The Euler-Kockel  effective potential, which is the first term of Euler-Heisenberg power series in $e^2/m_e^4$ is given by
\beqn
\Veff\simeq \frac{\alpha}{90\pi}\frac{e^2}{m_e^4}(4{\cal S}^2+7{\cal P}^2)+ \frac{2\alpha}{315\pi}\frac{e^4}{m_e^8}{\cal S}(8{\cal S}^2+13{\cal P}^2)+...
\eeqn
The common single power of $\alpha/\pi$ is due to the calculation being a one-loop evaluation of electron fluctuations.  For the coefficient functions, one finds
\begin{align}
\label{KEeps}
\frac{\partial\varepsilon}{\partial{\cal S}}
\simeq&-\frac{4}{45}\frac{\alpha}{\pi}\frac{e^2}{m_e^4}+\frac{32}{105}\frac{\alpha}{\pi}\frac{e^4}{m_e^8}{\cal S}+... \\
\label{KETmn}
\frac{1}{4}\frac{\partial\Tmm}{\partial{\cal S}}\simeq&\frac{4}{45}\frac{\alpha}{\pi}\frac{e^2}{m_e^4}{\cal S}-\frac{4}{315}\frac{\alpha}{\pi}\frac{e^4}{m_e^8}\left(24{\cal S}^2+13{\cal P}^2\right)+...
\end{align}

Let us compare Born Infeld  results \req{BIeps} and  \req{BITmn} to Euler-Kockel result \req{KEeps} and \req{KETmn}: we see that the QED effects are apparently more strongly suppressed because they arise from quantum corrections, signaled by the presence of $\alpha/\pi$.  With the smallness of $\alpha/\pi$, the Born-Infeld coefficients are $\sim 5000(m_e^4/M^4)$ times the QED coefficients and clearly this is excluded by precision QED tests. In fact the latest published constraint  on Born-Infeld $M\gtrsim 60~{\rm MeV}$~\cite{RafelskiBI2}, meaning $m_e^4/M^4\lesssim 6\:10^{-5}$, which assures that QED dominates  Born-Infeld corrections.  

\section{Physical effects of the nonlinear force}\label{sec:NLforce}

The force derived from the nonlinear action of QED or/and Born-Infeld type theory will originate in an increase in field energy when superposing fields. Thus in general, the force should act in such a way as to screen and reduce the total field strength.  In the case of a magnetic-only external field, this means the sign of the force depends on $\vec B_{\rm e}\cdot \vec H_{\rm p}$ displaying the relative orientation of the external field $\vec B_{\rm e}$ and probe magnetic field $\vec H_{\rm p}$.  For example, when the fields are aligned, $\vec B_{\rm e}\cdot \vec H_{\rm p}>0$ and the force acts to expel the probe from this domain. On the other hand, when the fields are anti-aligned, the force acts to pull probe deeper into the strong field domain to increase the screening.

The effect is manifested in the 4th (time-like) component of the force determines the change in energy of the probe current and is related the 3-force given shortly by the covariant requirement $u_{\mu}dp^{\mu}/d\tau=0$.  In the 0-component of \req{delta-4force}
we take the external electric field to be negligible, so the first term in \req{delta-4force} does not appear. With $F^{\rm e}_{\kappa\lambda}H_{\rm p}^{\kappa\lambda}=2\vec B_{\rm e}\cdot\vec H_{\rm p}$, we obtain
\begin{align}
\label{deltaf0}
\delta\! f^0=&\:{\cal S}_{\rm e}\:\partial_t\left(\frac{\partial\varepsilon}{\partial{\cal S}}2\vec B_{\rm e}\cdot\vec H_{\rm p}\right) 
-\partial_{t}\left(\frac{1}{4}\frac{\partial\Tmm}{\partial{\cal S}}2\vec B_{\rm e}\cdot \vec H_{\rm p}\right) \\   
\label{deltaf0v2}
=&\:4{\cal S}_{\rm e}\partial_t\left(\frac{\partial\varepsilon}{\partial{\cal S}}\vec B_{\rm e}\cdot\vec H_{\rm p}\right) +2\frac{\partial\varepsilon}{\partial{\cal S}}\vec B_{\rm e}\cdot\vec H_{\rm p}\partial_t{\cal S}_{\rm e}.
\end{align}
To obtain the second line, we use the identity
\beqn\label{Tmmepsreln}
-\frac{1}{4}\frac{\partial\Tmm}{\partial{\cal S}}={\cal S}\frac{\partial\varepsilon}{\partial{\cal S}}+{\cal P}\frac{\partial\varepsilon}{\partial{\cal P}}
\eeqn
which follows from the respective definitions Eqs.\:\eqref{epsdefn} \& \eqref{Tmmdefn}.  Again, since the external electric field is vanishing, the second term in this identity proportional to ${\cal P}=\vec E\cdot\vec B$ does not appear.

The 3-force corresponding to \req{deltaf0} is
\begin{align}
\delta\!\vec{f} 
\label{deltafk}
=&\:2\vec B_{\rm e}\cdot\vec H_{\rm p}\frac{\partial\varepsilon}{\partial{\cal S}}\vec B_{\rm e}\times(\vec\nabla\times\vec B_{\rm e}) \\ \nonumber
& +\left(|\vec B_{\rm e}|^2\vec\nabla-\vec B_{\rm e}(\vec B_{\rm e}\cdot\vec\nabla)\right)\left(2\vec B_{\rm e}\cdot\vec H_{\rm p}\frac{\partial\varepsilon}{\partial{\cal S}}\right)
 +\vec\nabla\left(2\vec B_{\rm e}\cdot\vec H_{\rm p}\frac{1}{4}\frac{\partial\Tmm}{\partial{\cal S}}\right) 
  \\ 
\label{deltafkv2}
=&\: 2\vec B_{\rm e}\cdot\vec H_{\rm p}\frac{\partial\varepsilon}{\partial{\cal S}}\vec B_{\rm e}\times(\vec\nabla\times\vec B_{\rm e}) 
-2\vec B_{\rm e}\cdot\vec H_{\rm p}\frac{\partial\varepsilon}{\partial{\cal S}}\vec\nabla{\cal S}_{\rm e}    \\ \nonumber
&+\left(\frac{|\vec B_{\rm e}|^2}{2}\vec\nabla -\vec B_{\rm e}\left(\vec B_{\rm e}\cdot\vec\nabla\right)\right)\left(2\vec B_{\rm e}\cdot\vec H_{\rm p}\frac{\partial\varepsilon}{\partial{\cal S}}\right)
\end{align}
The second equality is obtained again using \req{Tmmepsreln}.  The first term in \req{deltafkv2} corresponds to the first term in \req{delta-4force} and as seen in consideration of $\delta\! f^{0}$ does no work on the probe current.  It remains to be shown that the remaining two terms in \req{deltafkv2} are physically relevant. 

We consider two competing forces in an astrophysical plasma near a compact star: the gravity of the star and the internal energy (pressure) of the plasma.  For order of magnitude estimates of the force \req{deltafkv2}, we can use the Newtonian approximation for the star's gravity and replace the gradients by $1/L$ where $L$ is the compact star length scale, i.e. its radius $\sim 10~{\rm km}$.  Then, the ratio of radial forces is
\beqn\label{forceratio}
\frac{|\hat r\cdot\delta\!\vec{f}|}{|\hat r\cdot \vec f_{\rm grav}|} 
\simeq \frac{\partial\varepsilon}{\partial{\cal S}}\frac{|\vec B_{\rm e}|^3|\vec H_{\rm p}|}{L}\frac{L^2}{GM\rho}
=\frac{4\alpha}{45\pi}\frac{|\vec B_{\rm e}|^2}{|\vec B_c|^2}\frac{|\vec B_{\rm e}||\vec H_{\rm p}|}{\rho}\frac{L}{GM}
\eeqn
where $\rho$ is the mass density of the plasma $\rho\sim m_p 10^{11}~{\rm cm}^{-3}$ for a neutral astrophysical plasma. $G$ is the Newton gravitational constant and $M$ the mass of the star, so that $L/GM\sim 5-10$ for a compact star.  We have replaced $\partial\varepsilon/\partial{\cal S}$ by its QED value so that $|\vec B_c|$ is the QED ``critical'' field strength \req{Bc}. $|\vec H_{\rm p}|\gtrsim~$a few {\rm mG} according to estimates and models of accreting plasmas~\cite{Pessah}.  Putting in the numbers shows that the nonlinear electromagnetic force dominates gravity for $|\vec B_{\rm e}|/|\vec B_c|\gtrsim 10^{-7}$, confirming the previous analysis~\cite{Labun:2008gm}. 
We estimate the pressure $P$ in the plasma from the ideal gas law and compare it to the interaction energy density, which is essentially the force derived here without the gradients.  Thus, the ratio is
\beqn
\frac{|T^{00}_{\rm int}|}{P}
=\frac{\partial\varepsilon}{\partial{\cal S}}\frac{|\vec B_{\rm e}|^3|\vec H_{\rm p}|}{nT}
\eeqn
where $n$ is the number density $(\sim 10^{11}~{\rm cm}^{-3})$ and $T$ is the temperature, suggested by observations to be of order $10-100~{\rm eV}$~\cite{vanPeet}.  Since this differs from the previous estimate \req{forceratio} only by the factors $m_p/T\sim 10^{7}$ and $GM/L$, the immediate conclusion is the nonlinearity of the electromagnetic interaction may be more important than the plasma dynamics near to the surface of the star.  Note of course that a dipole field falls off rapidly as $|B_{\rm e}|\sim 1/r^3$ and these nonlinear field effects will be subdominant farther from the surface, say at 200 times the stellar radius where $|\vec B_{\rm e}|^3$ has fallen to $10^{-7}$ its value at the surface.

\section{Conclusions}

After reviewing the framework of nonlinear electromagnetism, we showed here how to calculate the force arising from the nonlinear field-field interaction, shown schematically in Fig.~\ref{FFforce}.  The procedure involves separately identifying the polarization of energy-momentum tensor related to presence of two components, here external field and particle field, so that we can account for the electromagnetic mass of particles. 

With only magnetic fields and Lorentz-Maxwell electrodynamics, the energy of charged particles near stellar objects is determined solely by gravity.  We have shown that  once Euler and Kockel~\cite{Euler:1935zz}  (effective) QED nonlinear electromagnetism is accounted for, magnetic fields can do work on charged particles and this effect beats out gravity for fields that are quite strong yet still very far from critical. This insight may have considerable physics impact, considering that there is no {\it a priori} limit to the magnetic field that a ferromagnetic star can source.  It is for this reason that exploration of the physical consequences arising in ultra strong magnetic fields remains a topic of current intense discussion relevant to extreme astrophysical events~\cite{1996ApJ...473..322T} and in various areas of elementary matter physics~\cite{Harding:2006qn,Colaiuda:2007br,Rabhi:2009ih,Ferrer:2005vd,Ferrer:2010wz,Sinha:2012cx}.

To close we note that the current report must be seen as a first step on the way to understand the classical charged particle dynamics in presence of ultra strong external magnetic fields and gravity. Beyond  the effects we considered here   we further expect radiation reaction to be relevant in  proximity of  the critical field strength where   acceleration is so strong that the radiation field impacts the source dynamics.

\begin{appendix}
\section{Finding the Lorentz force in NLEM}\label{app:LF}
In this appendix, we calculate the divergence of the electromagnetic energy-momentum tensor \req{Tmn-em}, 
\beqn\label{dTmn}
\partial_{\mu}\Tumn_{\rm e.m.} =(\partial_{\mu}\varepsilon)\Tumn_{\rm Max}+\varepsilon\partial_{\mu}\Tumn_{\rm Max}+\partial_{\mu}\gumn\Tmm/4
\eeqn
First, we write out the first two terms
\begin{align}
(\partial_{\mu}\varepsilon)&=\left(-\partial_{\mu}\frac{\partial{\cal L}}{\partial {\cal S}}\right)(\gumn{\cal S}-F^{\mu\kappa}F^{\nu}_{\phantom{\nu}\kappa})\label{deps}\\
\partial_{\mu}\Tumn_{\rm Max} &=\partial_{\mu}(\gumn{\cal S}-F^{\mu\kappa}F^{\nu}_{\phantom{\nu}\kappa})
\end{align}
with ${\cal L}=-{\cal S}+\Veff$ a shorthand for the total electromagnetic Lagrangian.  Then for the divergence of the trace $\Tmm$, we use the form given in \req{Tmmdefn},
\beqn\label{dTmm}
\partial_{\mu}\Tmm/4=+\left(\partial_{\mu}\frac{\partial{\cal L}}{\partial{\cal S}}\right){\cal S}+\left(\partial_{\mu}\frac{\partial{\cal L}}{\partial{\cal P}}\right){\cal P}
\eeqn
simplified by virtue of $d{\cal L}=(\partial{\cal L}/\partial{\cal S})d{\cal S}+(\partial{\cal L}/\partial{\cal P})d{\cal P}$.  Note that it does not matter whether one writes ${\cal L}$ or $\Veff$ in the variation because any terms linear in ${\cal S}$ manifestly cancel between the first and second terms of \req{Tmmdefn}.  The first term in \req{dTmm} then cancels with the first term in \req{deps}.  Next, we observe that
\beqn
-g^{\mu\nu} \mathcal{P}=   F^{\mu\kappa}\widetilde{F}_{\kappa}^{\phantom{\kappa}\nu}
\eeqn
and find for \req{dTmn}
\begin{align}
\partial_{\mu}\Tumn_{\rm e.m.}
=&\left(\partial_{\mu}\frac{\partial{\cal L}}{\partial {\cal S}}\right)F^{\mu\kappa}F^{\nu}_{\phantom{\nu}\kappa}-\frac{\partial{\cal L}}{\partial {\cal S}}\partial_{\mu}(\gumn{\cal S}-F^{\mu\kappa}F^{\nu}_{\phantom{\nu}\kappa})+\gumn\left(\partial_{\mu}\frac{\partial{\cal L}}{\partial{\cal P}}\right){\cal P}\\
\label{dTmnfinal}
=&\left(\left(\partial_{\mu}\frac{\partial{\cal L}}{\partial {\cal S}}\right)F^{\mu\kappa}+\frac{\partial{\cal L}}{\partial {\cal S}}\partial_{\mu}F^{\mu\kappa}+\left(\partial_{\mu}\frac{\partial{\cal L}}{\partial{\cal P}}\right)\widetilde{F}^{\mu\kappa}\right)F^{\nu}_{\phantom{\nu}\kappa}\\
&+\frac{\partial{\cal L}}{\partial {\cal S}}(F^{\mu\kappa}\partial_{\mu}F^{\nu}_{\phantom{\nu}\kappa}-\gumn\partial_{\mu}{\cal S}) \notag
\end{align}
Now using the definition of the displacement tensor \req{displacementdefn}, we recognize the first three terms in brackets as an expanded expression of the divergence of $-\Humn$.  Note that $\partial_{\mu}\widetilde{F}^{\mu\nu}$ vanishes according to the homogeneous Maxwell equation \req{homoMaxeqn}.  From \req{homoMaxeqn} one can also show that
\beqn\label{dSident}
\partial_{\mu}{\cal S}=\frac{1}{2}F^{\kappa\lambda}\partial_{\mu}F_{\kappa\lambda}=-\frac{1}{2}F^{\kappa\lambda}(\partial_{\kappa}F_{\lambda\mu}+\partial_{\lambda}F_{\mu\kappa})=-F^{\kappa\lambda}\partial_{\kappa}F_{\lambda\mu}
\eeqn
which means the final two terms in \req{dTmnfinal} cancel.  Thus, we obtain
\beqn
\partial_{\mu}\Tumn_{\rm e.m.}=j_{\mu}\Fumn
\eeqn
A slightly different proof, starting from a different but equivalent form of $\Tumn_{\rm e.m.}$ can be found in \S 6 of~\cite{BialynickiQED}.

\section{Lorentz force Correction derivation}\label{app:deltafderiv}
Considering the external field to provide the dominant part of the electromagnetic energy-momentum, we expand in the displacement tensor of the probe particle 
\beqn\label{Tmn-exp}
\Tumn_{\rm int}=\Tumn_{\rm e.m.}-\Tumn_{\rm e}-\Tumn_{\rm p} = 
\Tumn_{(1)}+\Tumn_{(2)}
\eeqn
\begin{subequations} \label{Tmnint-exp}
\begin{align}
\Tumn_{(1)} &= \left.\frac{\partial\Tumn_{\rm e.m.}}{\partial H^{\alpha\beta}}\right|_{\rm e} H_{\rm p}^{\alpha\beta} \\ 
\label{pp-int}
\Tumn_{(2)} &= \left.\frac{\partial^2\Tumn_{\rm e.m.}}{\partial H^{\alpha\beta}\partial H^{\kappa\lambda}}\right|_{\rm e}\!\!\!\frac{H_{\rm p}^{\kappa\lambda}H_{\rm p}^{\alpha\beta}}{2}-\Tumn_{\rm p}
\end{align}
\end{subequations}
with the subscript ${\bf e}$ reminding that derivatives are evaluated for the external field.  The zeroth order term is just $\Tumn_{\rm e}$, which is subtracted, and the energy-momentum of the probe particle is found at second order in $\Tumn_{(2)}$ and subtracted.  The expansion is in terms of the probe particle displacement tensor, because it is defined by the Maxwell equation with source \req{EMsource}.  

The derivatives with respect to $\Humn$ are related to derivatives with respect to $\Fumn$ by
\beqn\label{dFdH}
\frac{\partial F^{\alpha\beta}}{\partial \Humn}=
\frac{1}{2}(\delta^{\alpha}_{\mu}\delta^{\beta}_{\nu}-\delta^{\alpha}_{\nu}\delta^{\beta}_{\mu})+\frac{\partial^2 \Veff}{\partial \Fmn\partial F_{\alpha\beta}}+...
\eeqn
obtained from inverting \req{displacementdefn}.  The 1/2 is a symmetry factor arising from the equivalence under permutations of antisymmetric indices.   The two terms shown correspond to expanding up to order $e^4$ as needed in the present approximation.  Since $\Tumn$ is quadratic in $\Fumn$, expanding \req{dFdH} in powers of $e$ is the primary expansion involved in obtaining the result here. 

Using \req{Tmn-em}, the calculation of the tensor derivatives in \req{Tmnint-exp} is split into pieces:
\beqn
\frac{\partial \Tumn_{\rm Max}}{\partial H^{\kappa\lambda}}=\frac{\partial \Tumn_{\rm Max}}{\partial F^{\rho\sigma}}\frac{\partial F^{\rho\sigma}}{\partial H^{\kappa\lambda}}=\frac{\partial \Tumn_{\rm Max}}{\partial F^{\rho\sigma}}\left(\frac{1}{2}(\delta^{\rho}_{\kappa}\delta^{\sigma}_{\lambda}-\delta^{\rho}_{\lambda}\delta^{\rho}_{\kappa})+\frac{\partial^2 \Veff}{\partial F^{\rho\sigma}\partial F_{\kappa\lambda}}\right)
\eeqn
\beqn
\frac{\partial \varepsilon}{\partial H^{\kappa\lambda}}=\frac{\partial \varepsilon}{\partial F^{\rho\sigma}}\frac{\partial F^{\rho\sigma}}{\partial H^{\kappa\lambda}} 
\qquad\qquad 
\frac{\partial \Tmm}{\partial H^{\kappa\lambda}}=\frac{\partial \Tmm}{\partial F^{\rho\sigma}}\frac{\partial F^{\rho\sigma}}{\partial H^{\kappa\lambda}}
\eeqn
Because $\Veff$  and so $\varepsilon,\Tmm$ depend only on ${\cal S,P}$, derivatives with respect to $\Fumn$ can be split into partial derivatives
\beqn
\frac{\partial\:\cdot}{\partial\Fumn}=\frac{\partial\:\cdot}{\partial{\cal S}}F_{\mu\nu}+\frac{\partial\:\cdot}{\partial{\cal P}}\widetilde{F}_{\mu\nu}
\eeqn
We see then that $\partial\varepsilon/\partial{\cal S}, \partial\Tmm/\partial{\cal S}, \partial\varepsilon/\partial{\cal P}, \partial\Tmm/\partial{\cal P}$ are all order $e^4$, and the next-to-leading-order term in $\partial F/\partial H$ \req{dFdH} can be dropped.

Looking to the study of astrophysical magnetic fields, we specialize to the case $\vec E\cdot \vec B= {\cal P}=0$ for the external field.  Because $\Veff$ must be parity invariant, it contains only even powers of ${\cal P}^2$ and therefore the partial derivatives $\partial\varepsilon/\partial{\cal P}$ and $\partial\Tmm/\partial{\cal P}$ vanish being proportional to ${\cal P}$.  In this case then
\beqn\label{Tmnint}
\Tumn_{\rm int} = 
\Tumn_{\rm ep}-\Tumn_{\rm Max,e}\frac{\partial\varepsilon}{\partial{\cal S}}F^{\rm e}_{\alpha\beta}H_{\rm p}^{\alpha\beta}
+g^{\mu\nu}\frac{1}{4}\frac{\partial\Tmm}{\partial{\cal S}}F^{\rm e}_{\alpha\beta}H_{\rm p}^{\alpha\beta}
\eeqn
Here $\Tumn_{\rm Max,e}$ is the Maxwell energy-momentum tensor, compare \req{Tmn-em}, for the external field and the (linear) Maxwell-Lorentz interaction is separated:
\beqn\label{Tmnep}
\Tumn_{\rm ep} = -(F_{\rm e}^{\mu\kappa}H_{\rm p}^{\nu\lambda}+F_{\rm e}^{\nu\lambda}
H_{\rm p}^{\mu\kappa})g_{\kappa\lambda}+g^{\mu\nu}
\frac{1}{2}F^{\rm e}_{\alpha\beta}H_{\rm p}^{\alpha\beta}
\eeqn 
By going to the rest frame of the probe particle, it easy to show that this term produces the Lorentz force
\beqn\label{LF}
\partial_{\mu}\Tumn_{\rm ep}=j_{\mu}\Fumn_{\rm e}
\eeqn
Taking the negative divergence of \req{Tmnint} yields \req{delta-4force}.

Now, studying the second order term $\Tumn_{(2)}$ it turns out there are terms of order $e^4$.  Most can be seen to have form of the electromagnetic energy-momentum of the probe $\Tumn_{\rm p}$ (in its nonlinear form~\req{Tmn-em}).  One term has the form of a scalar modification to the Lorentz force, which is seen by differentiating $\varepsilon$ and $\Tumn_{\rm Max}$ each once.  However, we will consider the probe field to be much weaker than the external field, in which case these terms are all smaller than those calculated by a factor $|\vec H_{\rm p}|/|\vec B_{\rm e}|\ll 1$.  Incorporating these terms would provide the complete result for the 4-force at order $e^4$.

\end{appendix}
\vskip4mm
\noindent {\it Acknowledgments}  This work was supported in part by  the grant from the U.S. Department of Energy, DE-FG02-04ER41318. 


\end{document}